%% file: main.tex
\documentclass[sigconf]{acmart}

\AtBeginDocument{%
  }

\copyrightyear{2024} 
\acmYear{2024} 
\setcopyright{rightsretained} 
\acmConference[RecSys '24]{18th ACM Conference on Recommender Systems}{October 14--18, 2024}{Bari, Italy}
\acmBooktitle{18th ACM Conference on Recommender Systems (RecSys '24), October 14--18, 2024, Bari, Italy}\acmDOI{10.1145/3640457.3691718}
\acmISBN{979-8-4007-0505-2/24/10}

\begin{document}

%%
%% The "title" command has an optional parameter,
%% allowing the author to define a "short title" to be used in page headers.
\title[RecSys Algorithm Selection for Ranking Prediction on Implicit Feedback]{Recommender Systems Algorithm Selection for\\Ranking Prediction on Implicit Feedback Datasets}

%%
%% The "author" command and its associated commands are used to define
%% the authors and their affiliations.
%% Of note is the shared affiliation of the first two authors, and the
%% "authornote" and "authornotemark" commands
%% used to denote shared contribution to the research.
\author{Lukas Wegmeth}
\email{lukas.wegmeth@uni-siegen.de}
\orcid{0000-0001-8848-9434}
\affiliation{%
  \institution{Intelligent Systems Group\\University of Siegen}
  \city{Siegen}
  \country{Germany}
}
\author{Tobias Vente}
\email{tobias.vente@uni-siegen.de}
\orcid{0009-0003-8881-2379}
\affiliation{%
  \institution{Intelligent Systems Group\\University of Siegen}
  \city{Siegen}
  \country{Germany}
}
\author{Joeran Beel}
\email{joeran.beel@uni-siegen.de}
\orcid{0000-0002-4537-5573}
\affiliation{%
  \institution{Intelligent Systems Group\\University of Siegen}
  \city{Siegen}
  \country{Germany}
}

%%
%% By default, the full list of authors will be used in the page
%% headers. Often, this list is too long, and will overlap
%% other information printed in the page headers. This command allows
%% the author to define a more concise list
%% of authors' names for this purpose.
\renewcommand{\shortauthors}{Wegmeth et al.}

\input{content/abstract}

%%
%% The code below is generated by the tool at http://dl.acm.org/ccs.cfm.
%% Please copy and paste the code instead of the example below.
%%


%%
%% Keywords. The author(s) should pick words that accurately describe
%% the work being presented. Separate the keywords with commas.

%%
%% This command processes the author and affiliation and title
%% information and builds the first part of the formatted document.
\maketitle

\input{content/introduction}
\input{content/related}
\input{content/method}
\input{content/results}
\input{content/discussion}

%%
%% The next two lines define the bibliography style to be used, and
%% the bibliography file.
\bibliographystyle{ACM-Reference-Format}
\bibliography{sample-base}

\end{document}

%% file: content/abstract.tex
\begin{abstract}
The recommender systems algorithm selection problem for ranking prediction on implicit feedback datasets is under-explored.
Traditional approaches in recommender systems algorithm selection focus predominantly on rating prediction on explicit feedback datasets, leaving a research gap for ranking prediction on implicit feedback datasets.
Algorithm selection is a critical challenge for nearly every practitioner in recommender systems.
In this work, we take the first steps toward addressing this research gap.

%We employ meta-learning to solve the algorithm selection problem and explore the quality of tried-and-tested dataset meta-features.
%Additionally, we investigate the performance of automated machine-learning meta-models.
%The machine-learning community uses these techniques to address the algorithm selection problem.
%However, they have not yet been explored for solving the recommender systems algorithm selection problem.

We evaluate the NDCG@10 of 24 recommender systems algorithms, each with two hyperparameter configurations, on 72 recommender systems datasets.
We train four optimized machine-learning meta-models and one automated machine-learning meta-model with three different settings on the resulting meta-dataset.

Our results show that the predictions of all tested meta-models exhibit a median Spearman correlation ranging from 0.857 to 0.918 with the ground truth.
We show that the median Spearman correlation between meta-model predictions and the ground truth increases by an average of 0.124 when the meta-model is optimized to predict the ranking of algorithms instead of their performance.
Furthermore, in terms of predicting the best algorithm for an unknown dataset, we demonstrate that the best optimized traditional meta-model, e.g., XGBoost, achieves a recall of 48.6\%, outperforming the best tested automated machine learning meta-model, e.g., AutoGluon, which achieves a recall of 47.2\%.
\end{abstract}

%% file: content/introduction.tex
\section{Introduction}
The recommender systems algorithm selection problem for ranking prediction on implicit feedback datasets remains unsolved, and research on this topic is scarce.
Previous works on recommender systems algorithm selection focus on rating prediction and ranking prediction of explicit feedback datasets \cite{Beel2017-sf, DBLP:journals/corr/abs-1805-12118,DBLP:journals/corr/abs-1811-10369,collins2019first,DBLP:journals/corr/abs-2006-12328,DBLP:journals/corr/abs-2012-15151,10.1007/978-3-319-46227-1_25,10.1145/3109859.3109899,10.1007/978-3-319-67786-6_14,DBLP:journals/corr/abs-1807-09097,10.1145/3240323.3240378,DBLP:journals/corr/abs-1809-06120,CUNHA2018128}.
However, the recommender systems community has recently shifted its focus to solving ranking prediction on implicit feedback datasets.
Algorithm selection is a critical challenge for nearly every practitioner in recommender systems, underscoring its significant impact.

The algorithm selection problem is commonly defined as (automatically) finding the best algorithm for a given task and is a prominent problem in the machine-learning community \cite{8951014,8660744}.
Algorithm selection in machine learning and recommender systems is often solved with meta-learning techniques \cite{CUNHA2018128}.
Meta-learning here means to learn the relationship between dataset meta-features, also called dataset characteristics, and algorithm performance.

The machine-learning community boosted the performance of algorithm selection solutions with the introduction and development of automated machine-learning techniques \cite{erickson2020autogluontabularrobustaccurateautoml}.
However, to our knowledge, no works exist that explore the performance of automated machine-learning techniques on the recommender systems algorithm selection problem.

Recently, recommender systems research has shifted its focus toward solving ranking prediction tasks rather than rating prediction tasks.
That is, predicting the most relevant items to the user instead of predicting the rating a user would likely give an item.
The ranking prediction task was proposed over a decade ago \cite{4781121} and tackled in influential works already at least eight years ago \cite{10.1145/2959100.2959190}.

Similarly, the choice of recommender systems datasets has also changed over the years.
Traditionally, rating prediction was performed on explicit feedback datasets.
The predicted ratings were sometimes sorted and evaluated like a ranking prediction task.
However, with the shift to implicit feedback datasets in recommender systems practice, ranking prediction became the research focus.
Despite this, to our knowledge, there has been no research on the recommender systems algorithm selection problem for ranking prediction on implicit feedback datasets so far.

In explicit feedback datasets, the users provided an explicit weight of the interaction with an item to convey the strength of a like or dislike of the item.
In contrast, a weight is commonly absent in implicit feedback datasets, further constraining meta-features.
Steck \cite{10.1145/2507157.2507160} has addressed the contrast between the two tasks.
We think this warrants a study of whether the available evidence of recommender systems algorithm selection for rating and ranking prediction on explicit feedback applies to ranking prediction on implicit feedback datasets.

Given the introduced research gaps, we tackle the following research questions on the recommender systems algorithm selection problem for ranking prediction on implicit feedback datasets.
\begin{enumerate}
    \item[\textbf{RQ1:}]How effective are the established meta-features commonly used for solving the recommender systems algorithm selection problem for rating and ranking prediction on explicit feedback datasets when applied to ranking prediction for implicit feedback datasets?
    \item[\textbf{RQ2:}]How does the performance of automated machine-learning algorithms compare to traditional meta-learning algorithms in recommender systems algorithm selection for ranking prediction on implicit feedback datasets?
\end{enumerate}

To tackle the research questions, we develop a meta-dataset that includes the performance scores of 24 recommender systems algorithms, each with two hyperparameter configurations, on 72 recommender systems datasets.
For \textbf{RQ1}, we perform a literature review to find meta-features commonly extracted from explicit feedback datasets and understand whether they can be extracted from implicit feedback datasets.
We then train traditional meta-learning algorithms on our meta-dataset, evaluate their algorithm selection performance, and discuss the implications of the results.
For \textbf{RQ2}, we compare the algorithm selection performance of optimized traditional meta-learning algorithms versus automated machine-learning algorithms on our meta-dataset.
The results indicate whether automated machine-learning algorithms may be superior for solving the algorithm selection problem.

Our main contribution is the first analysis of the recommender systems algorithm selection performance for ranking prediction on implicit feedback datasets.
We compare traditional and automated machine learning meta-models using established meta-features for ranking prediction on implicit feedback datasets in recommender systems.
Furthermore, we are making our meta-dataset publicly available, which includes performance scores for 24 algorithms, each with two hyperparameter configurations, across 72 datasets, evaluated using three ranking metrics at five thresholds. 
The source code for reproducing all our results is available on GitHub\footnote{\url{https://code.isg.beel.org/RecSys-Algorithm-Selection-Ranking-Implicit-LBR}}.

%% file: content/related.tex
\section{Related Work}
Already over a decade ago, recommender systems researchers published works that correlate data characteristics, e.g., meta-features in the context of meta-learning algorithm selection, to algorithm performance \cite{10.1287/ijoc.1100.0385,10.1145/2245276.2245458,10.1145/2151163.2151166,10.1145/2365952.2366002,10.1145/2611040.2611054}.
Though having different objectives, they focus on understanding which data characteristics may predict the performance of a recommender systems algorithm for rating prediction.
All these works identify the common problem that no recommender systems algorithm is best for all datasets.

Following up on these works, roughly eight years ago, two groups of researchers, namely Beel \& Collins et al. and Cunha \& Soares et al., first analyzed the recommender systems algorithm selection problem as a meta-learning problem in a series of developing works (Beel \& Collins et al. \cite{Beel2017-sf, DBLP:journals/corr/abs-1805-12118,DBLP:journals/corr/abs-1811-10369,collins2019first,DBLP:journals/corr/abs-2006-12328,DBLP:journals/corr/abs-2012-15151,Wegmeth2022-bx}, Cunha \& Soares et al. \cite{10.1007/978-3-319-46227-1_25,10.1145/3109859.3109899,10.1007/978-3-319-67786-6_14,DBLP:journals/corr/abs-1807-09097,10.1145/3240323.3240378,DBLP:journals/corr/abs-1809-06120,CUNHA2018128}).
They provide concrete evidence of the performance of engineered meta-features and meta-learning algorithms on recommender systems algorithm selection in various domains.
In recent years other groups of researchers added new insights into the recommender systems algorithm selection problem \cite{10.1007/978-3-030-80568-5_39,9959945}.
Additionally, the Beel \& Kotthoff organized the AMIR workshop that focused on the topic \cite{Beel2019}. 
The shared focus of these works is understanding how to predict the best algorithm for rating or ranking prediction on explicit feedback datasets in recommender systems.

A few works have addressed recommender systems algorithm selection for ranking prediction tasks \cite{10.5555/3600270.3600589,CUNHA2018128,10.1145/3604915.3610656}.
McElfresh et al. \cite{10.5555/3600270.3600589} use meta-features that are only available in explicit feedback datasets.
They retrieve datasets that contain ratings, which they convert to weightless interactions for training recommender systems algorithms.
However, they extract meta-features that contain information about the interaction based on its rating before conversion.
Cunha et al. \cite{CUNHA2018128}, on the other hand, perform ranking prediction after predicting ratings by sorting the ratings.
Vente et al. \cite{10.1145/3604915.3610656} do not employ meta-learning but use the validation score during optimization. 
Our work differs from the others because we strictly focus on meta-learning with the constraints of implicit feedback datasets, where no rating information is available.

%% file: content/method.tex
\section{Method}
This section details our design decisions for the evaluation pipeline, specifically, which datasets and algorithms we choose for our meta-dataset, which meta-features we extract from the datasets, and which meta-learners we compare.

\subsection{Dataset Processing}
We retrieve 72 datasets\footnote{The interested reader may refer to our GitHub repository for a list of datasets.} from varying sources, shapes, and domains.
They include contain many popular recommendation datasets, e.g., variations of the MovieLens \cite{10.1145/2827872} and Amazon \cite{ni-etal-2019-justifying} datasets, and also less popular ones.
All datasets are designed explicitly for recommender systems applications.
For this first analysis, we constrain ourselves to datasets that contain up to one million interactions.

Since we focus on the algorithm selection problem for ranking prediction on implicit feedback datasets, we must convert explicit feedback datasets, e.g., ratings, to implicit feedback datasets, e.g., interactions.
We specifically address the problem of algorithm selection for implicit feedback datasets that are constrained by not having this type of weighting for interactions.
Therefore, we treat any rating as an interaction, as is commonly done.

We process every dataset using five-core pruning, which involves recursively removing users and items with fewer than five interactions. This helps to reduce noise and mitigates the impact of cold-start scenarios, as collaborative filtering algorithms struggle to learn from users and items without sufficient joint interactions.

Because an accurate estimation of algorithm performance is of utmost importance to the underlying algorithm selection problem, we employ five-fold cross-validation throughout the evaluation pipeline.
For example, we randomly split interactions per user into train and test sets at a ratio of 80\% to 20\%, ensuring that every interaction is tested once.
Our goal is to encompass the broadest range of data-constrained recommendation tasks.
Therefore, we choose not to apply a time-based split because datasets often do not contain timestamps.

\subsection{Meta-Features}
Literature on recommender systems meta-feature extraction primarily considers distribution meta-features \cite{10.5555/3600270.3600589,CUNHA2018128}.
In particular, counting the number of instances, features, labels, categories, etc., is straightforward.
For example, the number of users, items, and interactions, related information like data sparsity, and the minimum and maximum number of interactions of any user or on any item.

Extracting meta-features from the weightless interactions of implicit feedback datasets is more challenging than from explicit feedback datasets.
For example, when ratings are available, many meta-features use the rating of interactions, such as the mean rating, the histogram of ratings, and user and item bias.
We cannot use rating-based meta-features since we do not have ratings in implicit feedback datasets.
Interaction timestamps are also helpful for meta-feature extraction, e.g., the interaction time frequency, interaction history length, and the average time per user and item interaction.
However, we do not use time-based meta-features, as this would limit our algorithm selection findings to datasets with timestamps, which are sometimes absent in recommender systems datasets.

Therefore, we use the following meta-features in this paper: the number of users, the number of items, the number of interactions, the density of the user-item matrix, the ratio of users to items, the ratio of items to users, the highest number of ratings by a single user, the lowest number of ratings by a single user, the highest number of ratings on a single item, the lowest number of ratings on a single item, the mean number of ratings by each user, the mean number of ratings on each item.

\subsection{Algorithms}
We use 24 recommender systems algorithms\footnote{The interested reader may refer to our GitHub repository for a list of algorithms.} to present results for as many relevant algorithms as possible.
The algorithms are in various categories, e.g., neighborhood-based (User-based KNN, Item-based KNN), factorization-based (SVD, Implicit MF), deep learning (VAE, LightGCN), and popularity.
We further evaluate two hyperparameter configurations for each algorithm\footnote{Except Popularity and Random because they do not have hyperparameters.} to consider possible variations of algorithm performance due to hyperparameters.
This results in 46 different algorithm-hyperparameter combinations.
We use the algorithm implementations from RecBole \cite{recbole[1.2.0]}, LensKit \cite{10.1145/3340531.3412778}, and RecPack \cite{10.1145/3523227.3551472} to compare different libraries.

We calculate the number of recommender systems algorithm training procedures by multiplying the number of datasets by the number of data splits and algorithms-hyperparameter combinations.
In total, we train 16,560 recommender systems algorithms.
Due to this immense requirement, we constrain the training procedure to guarantee results after a particular time.
First, we limit ourselves to 8,280 GPU\footnote{On the OMNI cluster of the University of Siegen (AMD EPYC 7452, Tesla V100).} hours for training.
This results in precisely thirty minutes of training per algorithm, after which training is stopped, and the model at that time is used.
We acknowledge that half an hour of training may be limiting for specific algorithms. 
However, we guarantee that every algorithm produces a model in this time frame.
Finally, we choose three commonly used ranking metrics for recommender systems: \emph{nDCG}, \emph{Recall}, and \emph{Hit Rate}, and evaluate these metrics at multiple thresholds, e.g., 1, 3, 5, 10, and 20.

\subsection{Meta-Learner}
We use dataset meta-features as the input features for the meta-learning problem. 
The performance scores of recommender system algorithms on these datasets serve as the labels.
We aim to learn how dataset meta-features relate to algorithm performance to predict the best algorithm for a new dataset based solely on its meta-features.

Training the meta-learner is a machine learning problem, though under heavy constraints.
In this paper, we explore two different objectives for the meta-learning process: algorithm performance prediction and algorithm ranking prediction.
In algorithm performance prediction, we predict the performance of algorithms and then rank them.
In algorithm ranking prediction, we predict the ranking of algorithms directly.
The labels, e.g. the algorithm performance scores, are real numbers in performance prediction or integers in ranking prediction.
Therefore, for both objectives, we define the meta-learning problem as a regression problem.

We train one meta-learner per algorithm-hyperparameter combination, e.g., we pose the meta-learning problem as a single label-regression problem, where the label is the performance of an algorithm given a specific metric.
In this paper, we focus on NDCG@10.
This process is computationally expensive, but we expect more robust and fine-tuned models than posing the meta-learning problem, for example, as a multi-label regression problem.
Regardless, inference is fast because the meta-models are tiny.

For traditional meta-learning algorithms we employ the scikit-learn \cite{scikit-learn} implementation of Linear Regression, K Nearest Neighbors, and Random Forest, as well as XGBoost by DMLC \cite{Chen:2016:XST:2939672.2939785}.
All meta-learning algorithms are optimized with a grid search on a hyperparameter grid with more than 500 combinations.
To compare the traditional meta-learning algorithms with automated machine learning algorithms, we run AutoGluon \cite{erickson2020autogluontabularrobustaccurateautoml} with three settings: medium quality, best quality without bagging, and best quality with bagging, for up to twenty minutes each.

We perform a leave-one-out split for the evaluation of the meta-learning algorithms.
This means we train a model on all but one dataset and test it on the remaining dataset, repeating this for each dataset.
A leave-one-out split helps us to understand per dataset, whether the meta-learning is successful.
For most machine-learning tasks, a leave-one-out split would explode the training effort, as one model must be trained per instance.
However, due to the inherently small size of the meta-dataset, this is relatively inexpensive.
Multiplying the number of models, optimization objectives, and data splits yields 46,368 meta-models we train for this paper.

%% file: content/results.tex
\section{Results}
We present the comparison of the performance of traditional and automated machine-learning meta-models in the recommender systems algorithm selection problem for ranking prediction on implicit feedback datasets.
We focus on meta-models trained to predict the ranking or performance of recommender systems algorithms evaluated with the NDCG@10.

\begin{figure}[ht]   
    \centering
    \includegraphics[width=1\linewidth]{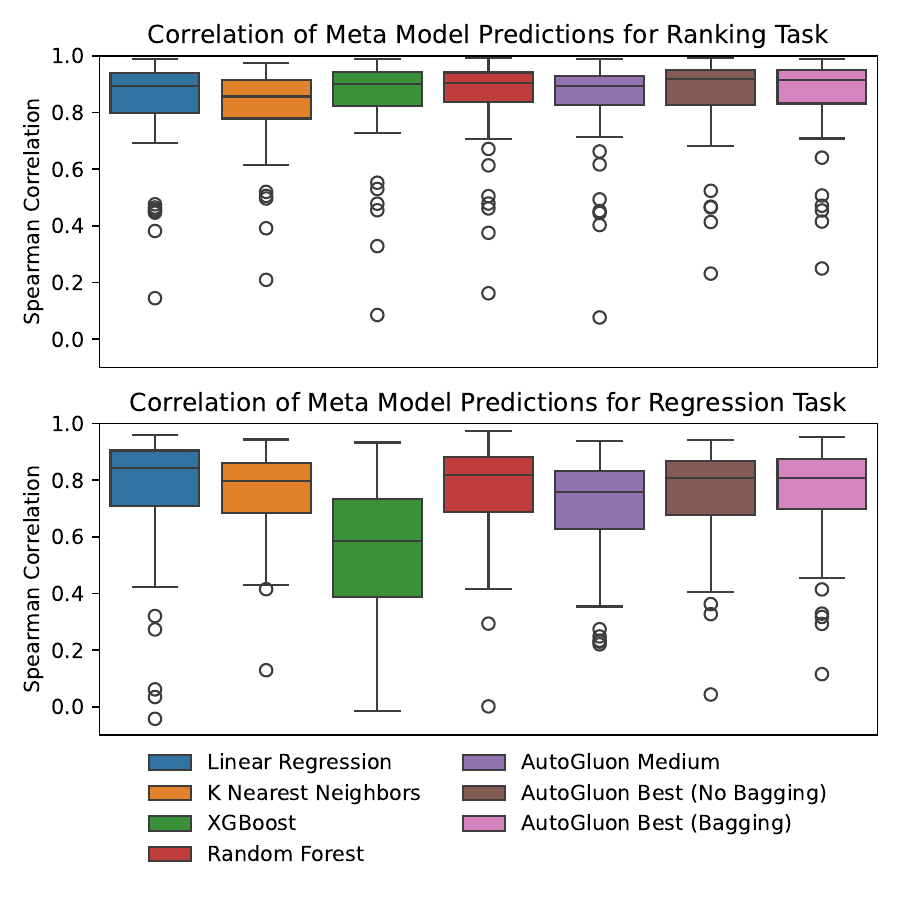}
    \caption{The Spearman correlation between meta-model predictions and ground truth (NDCG@10) per dataset. Each data point represents the correlation between predicted rankings (first plot) or performance (second plot) and the ground truth for a test dataset in a leave-one-out evaluation.}
    \label{corr}
\end{figure}

We begin by examining the Spearman correlation ($p < 0.05$) between the meta-model predictions and the ground-truth algorithm performances for each dataset, as shown in Figure \ref{corr}. 
Our analysis reveals a consistently high Spearman correlation across all meta-models. 
Notably, meta-models optimized for predicting algorithm rankings exhibit an average median Spearman correlation 0.124 points higher than those optimized for performance prediction.

Among the automated machine learning meta-models, AutoGluon Best (Bagging) achieves the highest Spearman correlation of 0.809. 
This is lower than the median Spearman correlation of 0.843 for the best traditional algorithm, Linear Regression, in performance prediction (see the second plot in Figure \ref{corr}). 
However, AutoGluon Best (No Bagging) outperforms the best traditional model, Random Forest, with a median Spearman correlation of 0.918 compared to 0.904 for ranking prediction (see the first plot in Figure \ref{corr}).

We also observe some outliers, where the meta-models struggle to learn effectively. 
Given that the ranking prediction objective consistently outperforms the performance prediction objective, we will focus on the ranking prediction objective moving forward.

\begin{figure}[ht]   
    \centering
    \includegraphics[width=1\linewidth]{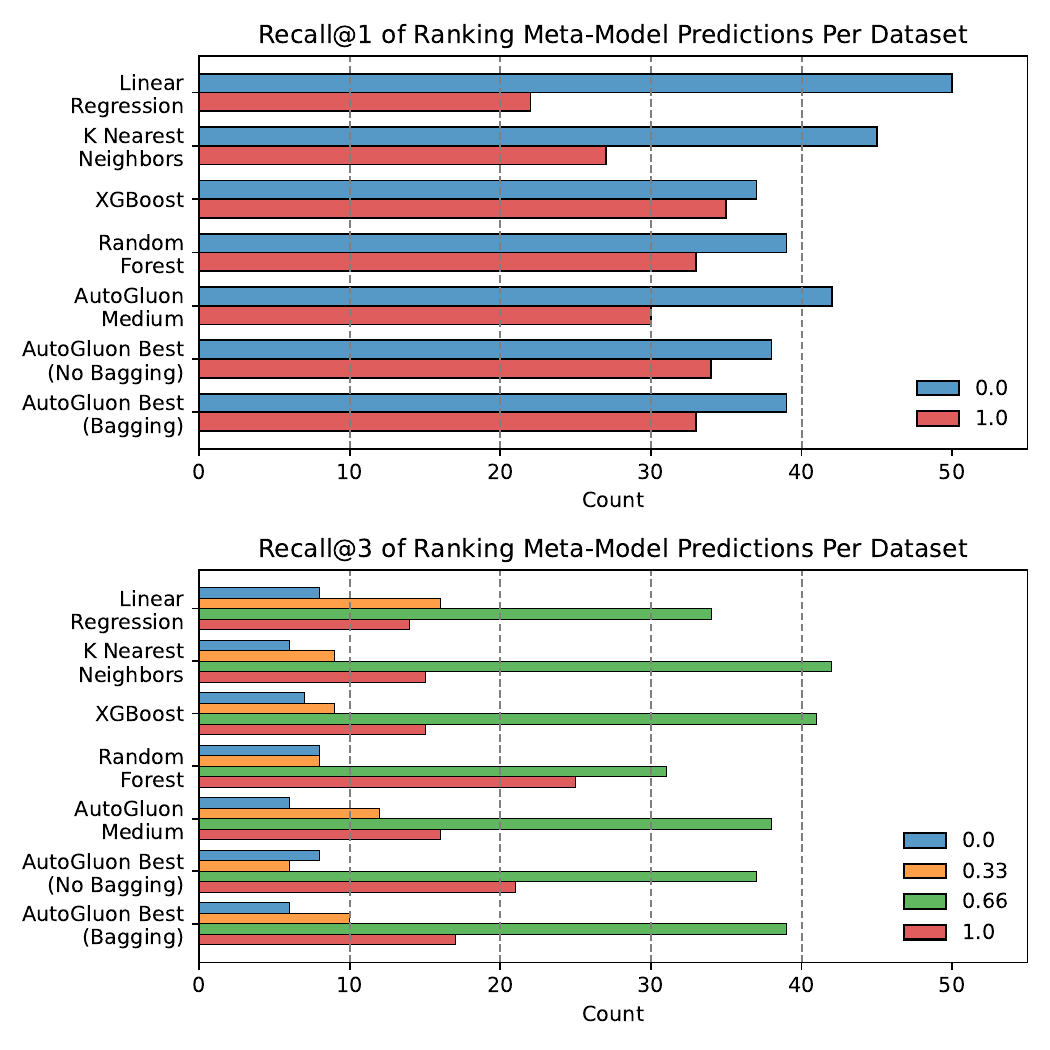}
    \caption{The Recall@1 and Recall@3 for the ranking objective meta-model predictions show the frequency of achieving the specified recall per dataset in a leave-one-out evaluation. For example, a Recall@1 score of 1 means the meta-model correctly identified the top algorithm. Each meta-model is evaluated on 72 datasets.}
    \label{recall}
\end{figure}

Most meta-models predict the best algorithm for nearly half of the datasets. 
Figure \ref{recall} illustrates this by showing the Recall of meta-models for the ranking prediction objective. 
The best meta-model for Recall@1 is the optimized XGBoost, with a score of 0.486, e.g., it predicts the best algorithm for 48.6\% of datasets. 
For Recall@3, all meta-models identify two of the top three algorithms in most cases. 
The best meta-model here is optimized Random Forest, with a Recall@3 of 0.669, predicting two of the top three algorithms for each dataset. 
Additionally, Random Forest predicts the top 3 algorithms for 34.7\% (25 of 72) of datasets. 
Traditional meta-models slightly outperform AutoGluon Best (No Bagging), which has a Recall@1 of 0.472 and a Recall@3 of 0.658. 
However, AutoGluon shows a higher Spearman correlation with the ground truth.

%% file: content/discussion.tex
\section{Discussion}
Answering \textbf{RQ1}, based on the presented results, we find that traditionally used meta-features are effective for predicting algorithm ranking.
Although we are unable to use many meta-features that consider ratings in original works on recommender systems algorithm selection, we show that even a limited set of meta-features leads to a high correlation between meta-model predictions and the ground truth.
We further demonstrate how we considerably improve the performance of the meta-models by optimizing them for predicting the ranking of algorithms instead of their performance.

Answering \textbf{RQ2}, based on the presented results, we find that the automated machine-learning meta-model AutuGluon has a higher correlation between the predicted algorithm ranking and ground truth than traditional optimized meta-models.
However, optimized traditional meta-models beat AutoGluon at predicting the best and the top three algorithms.
The performance difference between traditional models and AutoGluon is marginal but the training time for AutoGluon is higher.
Still, AutoGluon is easier to set up, requiring no parameter grid.

We are able to predict the best algorithm for 48.6\% of all datasets, regardless of size, domain, or algorithm category.
However, there is still much room for improvement, e.g., by extracting more complex meta-features, extending the meta-dataset, and improving meta-models.
In conclusion, we think that our results offer a positive outlook for the solution of the recommender systems algorithm selection problem for ranking prediction of implicit feedback datasets.